\documentclass[runningheads]{svmult}

\usepackage{makeidx}   
\usepackage{graphicx}  
\usepackage{subeqnar}  
\usepackage{multicol}  
\usepackage{physprbb}  
\makeindex             

\begin{document}

\title*{
Probing the Isotropy in the Sky Distribution 
of the Gamma-Ray Bursts
}
\toctitle{
Probing the Isotropy in the Sky Distribution 
of the Gamma-Ray Bursts
}
\titlerunning{Isotropy of Gamma-Ray Bursts}

\author{
Attila M\'esz\'aros\inst{1} \and
Zsolt Bagoly\inst{2} \and
Lajos~G.~Bal\'azs\inst{3} \and 
Istv\'an Horv\'ath\inst{4} \and
Roland Vavrek\inst{3}
}
\authorrunning{Attila M\'esz\'aros et al.}

\institute{
Astronomical Institute of the Charles University, V
Hole\v{s}ovi\v{c}k\'ach 2, CZ-180 00 Prague 8, Czech Republic
\and
Laboratory for Information Technology,
E\"{o}tv\"{o}s University, P\'azm\'any P\'eter
s\'et\'any 1/A, H-1518 Budapest, Hungary
\and
Konkoly Observatory, Box 67, H-1505 Budapest, Hungary
\and
Department of Physics, Bolyai Military University, 
Box 12, H-1456 Budapest, Hungary
}

\maketitle              

\begin{abstract}
The statistical tests - done by the authors - 
are surveyed, which verify the 
null-hypothesis of the intrinsic randomness in the angular distribution
of gamma-ray bursts collected at BATSE Catalog. The tests use
the counts-in-cells method, an
ana\-lysis of spherical harmonics, a test based on the two-point correlation 
function and a method based on multiscale methods. 
The tests suggest that the intermediate subclass of
gamma-ray bursts are distributed anisotropically.
\end{abstract}

\section{Introduction}

At the last years the authors carried out several statistical tests in order
to verify the isotropy of the angular distribution of the gamma-ray bursts
(GRBs) collected at BATSE catalog (\cite{mee00}). 
In this contribution we collect
the results of them; these results were partly published in several
articles (\cite{ba98}, \cite{ba99}, \cite{me00a}, \cite{me00b}, \cite{me00c}).

\section{Tests}

\subsection*{Spherical harmonics}

The key idea of this test is based on the fact that the sky-exposure function
of BATSE instrument 
is not depending on right ascension. Therefore in equatorial
coordinates the theoretically expected values of
spherical harmonics of the distribution of GRBs are zeros for any $ m
\neq 0$ term. Then these expectations are tested.

\subsection*{Counts-in-cells}

This is a simple statistical test. The idea is the following:
The sky is separated into equal areas, and
then, e.g., $\chi^2$ test is used to test the null hypothesis
of isotropy. The sky-exposure function can be
eliminated by the use of equatorial coordinates; then "effective" equal areas
are taken. For example, if the sky is separated into 8 equal areas,
then the boundaries are $\alpha = 0, 90, 180, 270$ degrees, 
 and $\delta = -30.8, +1.5, +33.6$ degrees
(instead of $\delta = -30, 0, +30$ degrees).

\subsection*{Two-point angular correlation function}

The key idea of this method is the following.
Having $N$ GRBs on sky we have $N(N-1)/$ angular 
distances among them. If $N$ GRBs are distributed randomly, then these
distances should be distributed randomly, too. Then the observed distances
are compared with the pseudo-randomly generated $N(N-1)/2$ distances 
coming from Monte Carlo simulations, which are provided in
accordance with the sky-exposure function.  Hence, the sky-exposure
function is eliminated by Monte Carlo simulations.

\subsection*{Multifractal analysis, minimal spanning tree, 
Voronoi tesselation}

For the detailed description of these three methods
see the contribution \cite{vav} in this Proceedings.

\section{Results}

The results of done tests for the three subclasses (\cite{kou},
\cite{hor}) of GRBs separately are collected at Table 1. 

\begin{table}
\caption{Survey of the results of the isotropy tests. The question
"Is the null hypothesis rejected?" is answered. When the answer is "Yes",
then the significance level of rejection is also given. We required a higher
than 95\% level.}
\begin{center}
\begin{tabular}{cccl}
\hline
short $T_{90} < 2 s\;\;\;\;$ & intermediate $2 s< T_{90} < 10 s\;\;\;\;\;$ 
& long
$T_{90} < 10 s\;\;\;$ & \\
\hline
No & Yes & No & Spherical \\
   & $>97\%$ &   & harmonics \\
\hline
No & Yes & No & Counts- \\
   & $>96.4\%$ &   & in-cells \\
\hline
Yes & Yes & Yes & Two-Point \\
$>99.2\%$   & $>99.8\%$ & $>99.8\%$  & Correlation \\
\hline
 No & Not & Not & Voronoi\\
  & done & done & tesselation \\
\hline
 No & Not & Not & Minimal\\
  & done & done & spanning tree \\
\hline
 Yes & Not & Not & Multifractal\\
$>99.9\%$  & done & done & analysis \\
\hline
\end{tabular}
\end{center}
\end{table}

\newpage

The done tests of isotropy
suggest the existence of {\bf ani\-sot\-ropy for the intermediate subclass}
on the confidence level $> 95\%$.

For the remaining two subclasses the situation is unclear;
there is no unambiguous rejection of isotropy for them yet on the
higher than 95\% confidence level. It can only be said that the
short subgroup is highly "suspicious".

\section{Conclusions}

The long GRBs seems to be distributed isotropically -
the positive result from two-point angular correlation function is probably
an unknown instrumental effect.

For the short GRBs the isotropy is not rejected yet on a
satisfactorily high confidence level, but 
there are indications for the anisotropy both
from the the multifractal analysis and also from the 
two-point angular correlation function. 
Add also that the statistical comparison of the 
short and the intermediate + long subgroups also suggests 
anisotropy here \cite{ba98}, \cite{ba99}. 
Simply the situation is highly "suspicious" here. Note still that
the shortest "tail" $T_{90} < 0.1$ s, which is doubtlessly
anisotropic \cite{cli00}, was not considered separately. 

The intermediate subclass \cite{hor} {\bf is anisotropic}; only the
concrete value of confidence level is a question - it "fluctuates"
between 96.4 - 99.9 \%.
The character of anisotropy of intermediate subclass is 
incomprehensible, because the "dimmer" half of this subsection
is more anisotropic \cite{me00b}. In addition, there is no concentration
toward the Galactic or Supergalactic planes.

This research was supported by Research Grant J13/98: 113200004 (A.M.),
by OTKA grants T024027 (L.G.B.), F029461 (I.H.) and T034549.

\end{document}